\documentclass[prl,twocolumn,amsmath,showpacs,floatfix]{revtex4}
\usepackage{amssymb}
\usepackage{graphicx}

\begin{document}
\input epsf

\title {Fast DNA translocation through a solid-state nanopore}

\author{Arnold J. Storm$^{1}$}
\author{Cornelis Storm$^{2}$}
\author{Jianghua Chen$^1$}
\author{Henny Zandbergen$^1$}
\author{Jean-Fran\c cois Joanny$^{2}$}
\author{Cees Dekker$^{1}$}

\affiliation{$^{1}$Kavli Institute of Nanoscience, Delft
University of Technology, Lorentzweg 1, 2628 CJ Delft, The Netherlands\\
$^{2}$Physicochimie Curie (CNRS-UMR168), Institut Curie Section de
Recherche, 26 Rue d'Ulm, 75248 Paris Cedex 05, France.}

\date{\today}

\begin{abstract}
We report translocation experiments on double-strand DNA through a
silicon oxide nanopore. Samples containing DNA fragments with
seven different lengths between 2000 to 96000 basepairs have been
electrophoretically driven through a 10 nm pore. We find a
power-law scaling of the translocation time versus length, with an
exponent of 1.26 $\pm$ 0.07. This behavior is qualitatively
different from the linear behavior observed in similar experiments
performed with protein pores. We address the observed nonlinear scaling in a theoretical model that describes experiments where hydrodynamic drag on the section of the polymer outside the pore is the dominant force counteracting the driving. We show that this is the case in
our experiments and derive a power-law scaling with an exponent of
1.18, in excellent agreement with our data.
\end{abstract}

\pacs{8714Gg,8715Tt}

\maketitle

Translocation of biopolymers such as polypeptides, DNA, and RNA is
an important process in biology. Transcribed mRNA molecules for
example are transported out of the nucleus through a nuclear pore
complex. Viral injection of DNA into a host cell is another
example. Translocation of DNA and RNA can be studied in vitro, as
demonstrated by Kasianowicz {\em et al.} \cite{kas96} using an
$\alpha$-hemolysin pore in a lipid membrane. By measuring the
ionic current through a voltage-biased nanopore, one can detect
individual single-strand molecules that are pulled through the
pore by the electric field. Li {\em et al.} \cite{li01,li03}
showed that solid-state nanopores can also be used for such
experiments. We report a set of experiments with silicon oxide
nanopores on double-strand DNA with various lengths. Surprisingly,
we find a nonlinear scaling between the translocation time $\tau$
and the polymer length $L_0$, in contrast to the linear behavior
observed for all experiments on $\alpha$-hemolysin
\cite{kas96,mel01}. In our experiments we find a clear power-law
relation $\tau \sim L_0^{1.26}$, for DNA fragments from 2000 to
96000 basepairs (bp). While a complete model for translocation
should in principle include hydrodynamic, steric, electrostatic
and entropic effects, we argue that in our experiments the
dominant contributions to the force balance come from
hydrodynamics and driving, and propose a simple model that
accurately reproduces the observed scaling.
\begin{figure}[!ht]
\begin{center}
\includegraphics[width=9cm]{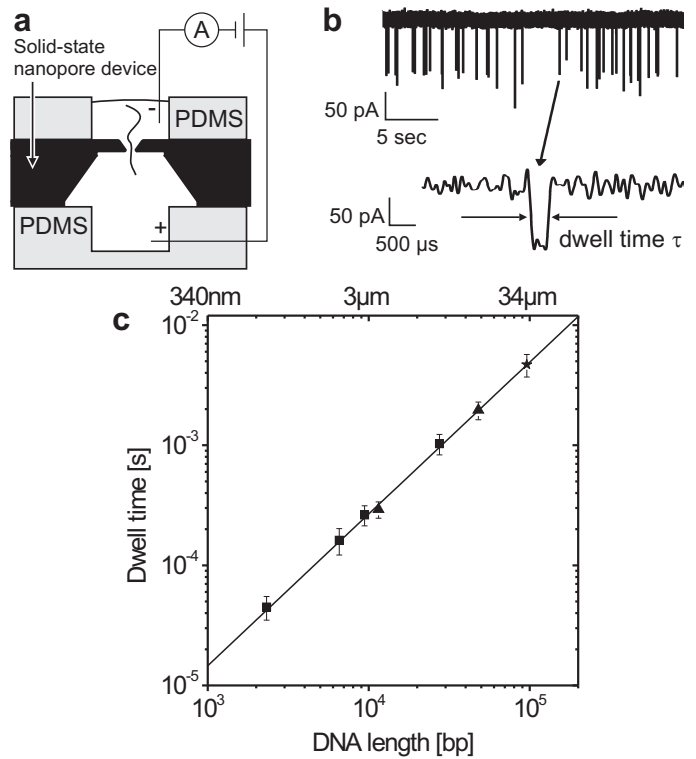}
\end{center}
\vspace{-.4cm} \caption[]{{\em (a)} Cross-sectional view of the
experimental setup (not to scale). A negatively-charged DNA
molecule is electrophoretically driven through a 10 nm aperture in
 a silicon device, which is located between two reservoirs kept at a potential difference.
Both reservoirs are filled with an aqueous buffer solution (1 M
KCl, 10 mM Tris-HCl pH 8.0, 1mM EDTA). {\em (b)} Measured ionic
current versus time, after the addition of 4 $\mu$m (11.5 kbp) DNA
to the top reservoir. An individual event is shown with increased
time resolution. {\em (c)} Dwell time vs. polymer length. The line
shows the result of a power-law fit to the data, with a best-fit
exponent of $\alpha=1.26 \pm 0.07$.}\label{expsystem}
\end{figure}

{\em Experimental results.} Figure \ref{expsystem}{\em (a)} shows
the experimental layout for translocation studies. At the heart of
the setup is a solid-state nanopore device, fabricated by
shrinking a 20-50 nm pore in silicon oxide in a transmission
electron microscope to a final size of 10 nm \cite{sto03}. The
nanopore is situated in an insulating membrane which separates two
macroscopic reservoirs filled with an aqueous buffer solution.
When a voltage bias is applied over the membrane in the presence
of DNA molecules in the negative compartment, the DNA is
electrophoretically drawn through the pore due to its negative
charge. The detection technique is simple and elegant: A polymer
traversing the pore lowers the amount of conducting solution
present inside the pore and thereby reduces the ionic conductivity
between the reservoirs. Passing molecules are thus detected as
short dips in the ionic current, which is induced by the
externally applied voltage (see Fig.~\ref{expsystem}{\em (b)}).
Analogous to Li {\em et al.} \cite{li03}, we find that the
molecules with a diameter of about 2 nm can pass the 10 nm pore
either in a linear or in a folded fashion. Using an event sorting
algorithm discussed elsewhere \cite{sto04}, we analyze only the
linear unfolded translocation events in this work.
Fig.~\ref{expsystem}{\em (b)} shows an example of such a linear
translocation event, detected with 11.5 kbp linear DNA. The width
of the dip is interpreted as the duration of the translocation. We
performed three experiments, all at room temperature: One on
linear 11.5 kbp DNA, one on linear 48 kbp $\lambda$-DNA (here we
detected both individual molecules and dimers of two molecules
bound together with their complementary sticky ends), and one on a
mixture that contains 27491 bp, 9416 bp, 6557 bp, 2322 bp, and
2027 bp fragments. In the last experiment, the length difference
between the 2322 and 2027 bp fragments could not be resolved
experimentally. The durations of individual linear events were
collected in a histogram for each experiment. Figure
\ref{expsystem}{\em (c)} shows the dwell time (determined as the
peak position in these histograms) versus polymer length for all 7
DNA fragments. The full width at half maximum of the peaks are
taken as error bars. We find a clear power-law scaling of the
dwell-time $\tau$ with the length $L_0$, $\tau \sim L_0^{\alpha}$,
with a value of 1.26 $\pm$ 0.07 for the exponent $\alpha$. These
experiments are discussed in more detail in \cite{sto04}. It is
interesting to note that our results are in good agreement with
those reported by Li {\em et al.} \cite{li03}. They have studied
the translocation of 3000 bp and 10000 bp DNA through a 10 nm
silicon nitride nanopore, and find translocation times of 100
$\mu$s and 400 $\mu$s respectively.

The translocation process consists of two separate stages. First,
there is the capture stage. A DNA molecule initially in solution
in the negative reservoir has to come close enough to the pore to
experience the electrostatic force and get pulled in. We assume
that the reservoirs are good ionic conductors, and the driving
force is only felt in the direct vicinity of the pore. Capture is
thus a stochastic process, since the pore has to be reached by
diffusion. In this work, we focus on the second stage, where the
DNA passes the pore until it has reached the other side. We assume
that one end of the DNA has entered the pore and calculate the
time required for complete translocation.

{\em Slow vs. fast translocations.} We now address the dependence
of this duration on the length of the polymer. To this end, we
consider a linear polymer consisting of $N$ monomers, each of
which has a Kuhn length $b$. This polymer is partially threaded
through a narrow pore. Time $t=0$ sets the moment of initial
capture. We will let $L(t)$ denote the contour length of the
untranslocated part of the polymer, so that $L(0)=Nb\equiv L_0$.
The dwell time $\tau$ is therefore determined by $L(\tau)=0$. A
second time scale in the problem is the characteristic relaxation
time scale of the translocating polymer. This Zimm time
\cite{doi},  given approximately by
\begin{equation}
t_Z \approx 0.4\,\frac{\eta R_g^3}{k_B T}\, ,
\end{equation}
can be considered an upper bound on the time it takes the polymer
to relax to an entropically and sterically favored configuration.
In this expression, $\eta$ is the solvent viscosity and $R_g$ is
the radius of gyration of the polymer. This is the
radius of the typical blob-like configuration that the polymer
will assume in a good solvent, and it scales with the polymer
length as
\begin{equation}
R_g \sim L_0^\nu \, ,\label{rgvsL}
\end{equation}
which defines the swelling exponent $\nu$. It depends on the
dimensionality of the system, and theoretically a value of 0.588
is found for self-avoiding polymers in a good solvent
\cite{zinnjustin}. Smith {\em et al.} \cite{smi96} have measured
the diffusion constant $D$ for stained DNA molecules with lengths
ranging from 4.3 kbp to 300 kbp. They report a scaling with length
$L_0$ as $D \sim L_0^{- \nu}$ with $\nu = 0.611 \pm 0.016$, and
conclude that Flory scaling is appropriate for DNA molecules
longer than about 4 kbp. At room temperature, the measured
velocity is about 0.8 $\mu$s per base or slower \cite{mel00}. A
100-base, single-stranded DNA fragment therefore takes around 80
$\mu$s to fully translocate. When we compare this to the Zimm time
for the same polymer fragment, about 0.2 $\mu$s, we see that
relaxation is much quicker than the translocation. We will call
such events, for which $\tau \gg t_Z$, {\em slow} translocations.
Lubensky and Nelson \cite{lub99} have argued that for
single-stranded DNA and RNA through $\alpha$-hemolysin, the
criterion for slow translocation is indeed satisfied for polymer
lengths up to hundreds of nucleotides. They show that the Zimm
time for a polynucleotide of roughly 300 bases is comparable to
the translocation time per nucleotide.

The criterion for slow translocation is evidently not met in our
experiments on solid-state nanopores. A full $\lambda$-phage
genome (48.5 kbp, or 16.5 $\mu$m of double-stranded DNA) is found
to take only around 2 ms to traverse a 10 nm SiO$_2$ pore. The
Zimm time for this molecule, in comparison, is about 700 ms,
clearly much longer than the translocation time. Even the
translocation of the shortest molecules studied in these assays
(2000 bp) can be considered fast, with a translocation time of
about 50~$\mu$s and a Zimm time of just over 2.3 ms. We therefore
refer to this second regime, where $\tau \ll t_Z$, as {\em fast}
translocations. We should point out that an important reason for
the fastness of our system is the fact that we use double-stranded
DNA, which has a much larger persistence length than
single-stranded DNA and consequently has a long relaxation time.

\begin{figure}
  \begin{center}
  \includegraphics[width=7cm]{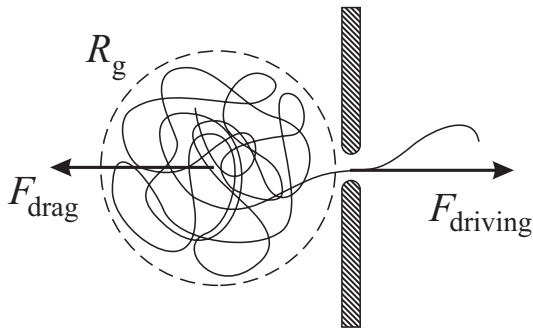}\\
  \end{center}
  \caption{At time $t$ the DNA on the left side of the pore has a radius of gyration of
$R_g$, indicated by the dashed line. The balance between the two
dominating forces determines the dynamics: A driving force that
locally pulls the DNA through the pore and a viscous drag force
that acts on the entire DNA blob.}\label{model}
\end{figure}

{\em Model.} Let us estimate the magnitudes of the possibly
relevant forces, following Lubensky and Nelson \cite{lub99}.
First, consider the driving. As stated, a potential difference
across the pore exerts a highly localized force on the negatively
charged DNA molecule. We assume the potential drop to occur
entirely inside the pore, and therefore only the part of the
polymer inside experiences the driving force. This force can then
be estimated as $F_\text{driving}= 2 eV/a$, where $e$ is the
elementary charge, $V$ is the potential difference and $a=0.34$ nm
is the spacing between nucleotides. A bias voltage of 120 mV, as
is typically used in experiments, thus produces a force of around
110 pN. This value is an upper bound of the actual force, as
screening effects may greatly reduce the effective charge on the
DNA, and thereby the driving force. Simulations of Manning
condensation on double-stranded DNA yield charge reduction values
between 53\% and 85\% \cite{fraction1,fraction2}. Barring complete
screening however, we consider our DNA translocations to be
strongly driven and this justifies ignoring diffusive
contributions.

In the absence of specific DNA-pore interactions, the viscous drag
per unit length in the pore can be estimated as $2 \pi \eta
rv/(R-r)$, where $R$ is the pore radius, $r$ is the polymer
cross-sectional area, $\eta$ the solvent viscosity and $v$ is the
linear velocity of the polymer inside the pore. Substituting
typical values ($\eta=1.10^{-3}$ Pa$\cdot$s, $r=2$ nm, $v=10$
mm/s, $R=10$ nm and a pore depth $\ell_{\text{pore}}$ of 20 nm) we
can estimate this drag force to be around 0.3 pN, decidedly
smaller than the driving force. We feel this constitutes an
essential difference between solid-state pores and protein pores:
In sufficiently shallow solid state pores the effect of friction
inside the pore is negligible.

Finally we estimate the hydrodynamic drag on the untranslocated
part of the polymer, outside the pore. To this end, we approximate
the untranslocated part as a sphere of radius $R_g$ (see
Fig.~\ref{model}). As the polymer threads through the pore, the
center of mass of this sphere moves towards the pore at a velocity
${\rm d}R_g/{\rm d}t$. Assuming Zimm dynamics (and thus that the
solvent inside the coil moves with the polymer), the coil
experiences a Stokes drag force of $6 \pi \eta R v=6 \pi\eta R_g
{\rm d}R_g/{\rm d}t$, which for typical parameters yields a drag
force of about 24 pN. This assumption is justified by experiments
by Smith {\em et al.} \cite{smi96}, who found clear evidence for
Zimm dynamics for DNA longer than 4.3 kbp. Clearly, in this case
the hydrodynamic friction on the part of the polymer outside the
pore is the dominant force counteracting the driving force. We
therefore choose to model fast translocation dynamics as
determined only by the cumulative effect of driving at the pore
and hydrodynamic friction outside.

Figure \ref{model} depicts the simplified system we consider. The
part of the polymer inside the pore experiences a driving force to
the right, while the length of polymer before the pore is coiled
up. The pore is sufficiently small to allow only linear ({\em
i.e.} unfolded) passage of a single molecule at a time.

As the polymer is pulled through the pore the blob before the
entrance shrinks in size, and thus its center of mass moves
towards the pore with a velocity
\begin{equation}
v_{\text{blob}} = \dot{R_g} \sim L^{\nu-1}\dot{L} \, ,
\end{equation}
where the dot denotes a time derivative. Motivated by our
consideration of the relative magnitudes of the counteracting
forces, we propose the principal effect of hydrodynamics is to
resist motion with a Stokes drag on the DNA coil that is
proportional to its radius times the velocity,
\begin{eqnarray}
    F_{\text{drag}} &\sim& Rv_{\text{blob}} \sim R_g \dot{R_g} \nonumber \\
             &\sim& L^{2\nu-1}\dot{L} \label{FdragvsL}\, .
\end{eqnarray}

Force balance must be met at all times, and since there are only
two major forces the driving force should balance the hydrodynamic
friction: {$F_{\text{drag}}=-F_{\text{driving}}$}. As the driving
force is constant during the whole process, the same holds for
$F_{\text{drag}}$. Thus we can extract the linear velocity
$v_{\text{lin}} = -\dot{L}$ of the DNA inside the pore from
Eq.~(\ref{FdragvsL}):
\begin{equation}
    v_{\text{lin}} = -\dot{L} \sim -L^{1-2\nu}\label{v(L)},
\end{equation}
which allows us to obtain the dwell time $\tau$ by integration,
\begin{equation}\label{result}
\tau = \int_0^{\tau}\!{\rm
d}t=\int_{L_0}^0\!\!v^{-1}_{\text{lin}}(L)\,{\rm d}L\sim
L_0^{2\nu}.\nonumber
\end{equation}

On the basis of this model we thus predict a powerlaw relation
between the dwell time $\tau$ and the contour length $L_0$. Taking
the theoretical value of 0.588 for the Flory exponent $\nu$ we
find an exponent of $\alpha=2\nu=1.18$ for this model. If we take
the experimentally obtained value for $\nu$ of 0.61 \cite{smi96},
we find $\alpha = 1.22$ - in excellent agreement with our
experiments, where we find power law scaling with an exponent of
1.26 $\pm$ 0.07.

{\em Scaling regimes for translocation.} General considerations
along the lines of the argument presented in the preceding
sections can be used to qualitatively understand the various
regimes of polymer translocation. Firstly, it is important to
determine the dominant contribution to the friction. In most
cases, it suffices to compare the pore friction
$F_{\text{pore}}=\xi_{\text{eff}}\, v_{\text{lin}}$ (with
$\xi_{\text{eff}}$ equal to $2\pi \eta \ell_{\text{pore}}
\,r/(R-r)$ in the absence of specific interactions) to the Stokes
drag on the coil $6 \pi \eta R_g \dot R_g$. If the pore friction
dominates, force balance with respect to the constant driving
force implies that the translocation time scales linearly with the
polymer's length $\tau \sim L_0$. A possible reason for a large
pore friction could be the presence of specific interactions, but
because of the geometric factor in the effective friction constant
$\xi_{\text{eff}}$ the {\em shape} of the pore could also lead to
pore friction dominated translocation. Such linear dependence of
$\tau$ on the length for single-stranded DNA ranging from 12 to
400 bases has been reported experimentally by Kasianowicz
\cite{kas96} and Meller \cite{mel01}. For the $\alpha$-hemolysin
pore they used it is indeed speculated that significant specific
interactions with the passing DNA occur.

When the Stokes drag dominates one can derive, without any
assumptions on the polymer statistics, that $\tau\sim R_g^2$.
Depending on the length of the polymer different regimes are thus
obtained: when the polymer is short compared to its persistence
length $R_g \sim L_0$, and we find that $\tau\sim L_0^2$. For
polymers of intermediate length the radius of gyration follows the
scaling for a Gaussian chain, $R_g\sim L_0^{1/2}$, and
consequently the translocation time is predicted once again to
scale linearly with length (note, however, that this is a
qualitatively different regime than the pore-friction dominated
regime identified before). For long polymers (such as those
considered in the preceding sections) we have shown that $\tau\sim
L_0^{2 \nu}$.

So far we have presented scaling arguments assuming Zimm (non-free
draining) dynamics. Kantor and Kardar have identified and
numerically confirmed yet another regime \cite{kan03} where
$\alpha=\nu\!+\!1$. This behavior can be understood assuming Rouse
dynamics (stationary solvent). In this case, the hydrodynamic drag
would be given by $F_{\text{drag}}\sim L v_{\text{blob}}$, and one
does indeed recover their scaling law $\tau \sim L_0^{1+\nu}$. We
speculate that this regime might be observable for semidilute
solutions close to $c^\star$.

In all cases considered in this section, one can independently
determine whether or not the polymer is frozen in its
configuration by comparing the Zimm- and translocation timescales.
We do not expect this to affect the scaling behavior in the
various regimes, but it {\em will} affect the prefactors.

{\em Concluding remarks.} We have obtained a  simple and elegant
model description that appears to describe our data well. There
are however several effects we neglect but which could have an
additional influence on the process that we consider. For
instance, we ignore any friction experienced by the DNA that has
already passed the pore. We also expect that an electro-osmotic
flow is generated inside the pore. This effect is caused by an
electrophoretic force on the ions screening the charge on the
surface of our pore. As silicon oxide is known to be negatively
charged in water, there is a surplus of positive ions near the
surface. These positive ions generate a flow of water inside the
pore, slowing down the DNA that moves in the other direction.
While we have not explored the consequences of these possibilities
the observed agreement between theory and experiment suggests that
at least for the fast polymer translocations considered here,
hydrodynamic drag does indeed dominate the dynamics.
Identification and understanding of the dominant effects in
polymer translocation through nanopores is relevant not only for
biological processes, but also for potential analytical techniques
based on nanopores. Rapid oligonucleotide discrimination on the
single-molecule level has been demonstrated with $\alpha$
hemolysin \cite{mel00}, and more recently solid-state nanopores
were used to study folding effects in double-stranded DNA
molecules \cite{li03,sto04}. Future applications of this technique
may include DNA size determination, haplotyping and sequencing.

{\em Acknowledgements.} We would like to thank the groups of
D.~Branton and J.~A.~Golovchenko at Harvard University for their
hospitality and advice on nanopore experiments. We also thank
Nynke Dekker, Sean Ling, John van Noort and Peter Veenhuizen for
their support and discussions. This research was financially
supported by the Dutch Foundation for Fundamental Research on
Matter (FOM). C.~S. acknowledges funding from the European PHYNECS
research network. J.~H.~C. is grateful to the Netherlands
Institute for Metals Research (NIMR) for financial support under
Project MC4.98047.

\end{document}